\documentclass{emulateapj}

\shorttitle{Electron Acceleration by Multi-Island Coalescence}
\shortauthors{Oka et al.}

\begin{document}


\title{Electron Acceleration by Multi-Island Coalescence}


\author{M. Oka\altaffilmark{1}, T.-D. Phan\altaffilmark{1}, S. Krucker\altaffilmark{1}, M. Fujimoto\altaffilmark{2}, I. Shinohara\altaffilmark{2}}

\altaffiltext{1}{Space Sciences Laboratory, University of California Berkeley}
\altaffiltext{2}{Institute of Space and Astronautical Science,\\Japan Aerospace Exploration Agency}

%
%
%

%


\begin{abstract}
Energetic electrons of up to tens of MeV are created during explosive phenomena in the solar corona. 
While many theoretical models consider magnetic reconnection as a possible way of generating energetic electrons, the precise roles of magnetic reconnection during acceleration and heating of electrons still remain unclear. Here we show from 2D particle-in-cell simulations that coalescence of magnetic islands that naturally form as a consequence of tearing mode instability and associated magnetic reconnection leads to efficient energization of electrons. The key process is the secondary magnetic reconnection at the merging points, or the `anti-reconnection', which is, in a sense, driven by the converging outflows from the initial magnetic reconnection regions.  By following the trajectories of the most energetic electrons, we found a variety of different acceleration mechanisms but the energization at the anti-reconnection is found to be the most important process. We discuss possible applications to the energetic electrons observed in the solar flares. We anticipate our results to be a starting point for more sophisticated models of particle acceleration during the explosive energy release phenomena.
\end{abstract}





\section{Introduction}

A solar flare is an explosive energy release phenomena on the sun 
and a large fraction of the released energy appears to go to high energy, often non-thermal, particles both ions and electrons \citep[][and references therein]{lin03}. The particle energy reaches tens of GeV for ions and tens of MeV for electrons. The mechanism of producing such energetic particles is much less understood compared to the energy release mechanism. Because magnetic reconnection is a possible mechanism of the energy release process, particle acceleration may also occur through magnetic reconnection, although difficulties remain when trying to interpret observations \citep[][and references therein]{miller97, krucker08, benz08}. In order to explain observations, many theoretical ideas have been proposed \citep[][and references therein]{aschwanden02}. While some theories utilize fast/slow mode shocks as well as electromagnetic waves of various scales - sometimes in a stochastic manner - many theories consider magnetic reconnection.  In this paper, we also assume {\it a priori} that magnetic reconnection plays a role for particle acceleration and explore possible mechanism(s) of particle acceleration in association with magnetic reconnection.

In general, a test particle approach had been used to explore particle acceleration by magnetic reconnection. While some studies solved particle motion analytically \citep[e.g.][]{litvinenko96}, a majority of studies performed test-particle simulations under model electromagnetic fields \citep[e.g.][]{kliem94, hannah06, onofri06} or time varying fields generated by MHD simulations \citep{sato82,scholer87, ambrosiano88}. A self-consistent, particle-in-cell (PIC) simulations are now widely used to study the detailed process of electron energization \citep[e.g.][]{hoshino01,hoshino05, drake05, pritchett06, drake06, karlicky07, pritchett08, wan08, shinohara09}. PIC simulations have an advantage of being able to resolve the inner structure of the X-line, the so-called diffusion region. It is a scientific challenge to understand particle acceleration by magnetic reconnection that involves multi-scale. 

For convenience, we categorize various theories of particle acceleration associated with magnetic reconnection into two different groups: {\it X-type} acceleration and {\it O-type} acceleration. The {\it X-type} acceleration takes place at and around the X-line of magnetic reconnection or the diffusion region. Particles are unmagnetized at the X-lines and can be directly accelerated by the electric field \cite[e.g.][]{sato82}. In the immediate downstream of the X-line are the regions with magnetic field gradient where particles further gain energy while drifting along the current sheet \citep[e.g.][]{scholer87,kliem94,hoshino01}. 
More recently, it was found that the in-plane, polarization electric field in the diffusion region generated by the charge separation between ions and electrons plays an important role \citep{hoshino05}. The force by the polarization electric field can be balanced by the Lorentz force so that electrons are accelerated efficiently by the reconnection electric field while being trapped within the current sheet boundary.  Because of the trapping effect, the process was named `surfing' mechanism. 



The {\it O-type} acceleration takes advantage of the closed geometry of field lines in a magnetic island. In many cases, a magnetic island is bounded by two X-lines at each end. Therefore, if particles are trapped in a magnetic island, they can continue gaining energy by repetitive crossings of the gradient region, although the reconnection electric field is relatively weak inside magnetic islands \citep{stern79,scholer87,kliem94}. In order to compensate for this weak electric field, a recent model has been developed that takes into account the dynamical, contracting motion of islands \citep{drake06}. The time-dependent model is analogous to the energy increase of a ball reflecting between two converging walls, namely the first order `Fermi' process. 


\begin{figure}
\epsscale{1.0}
\plotone{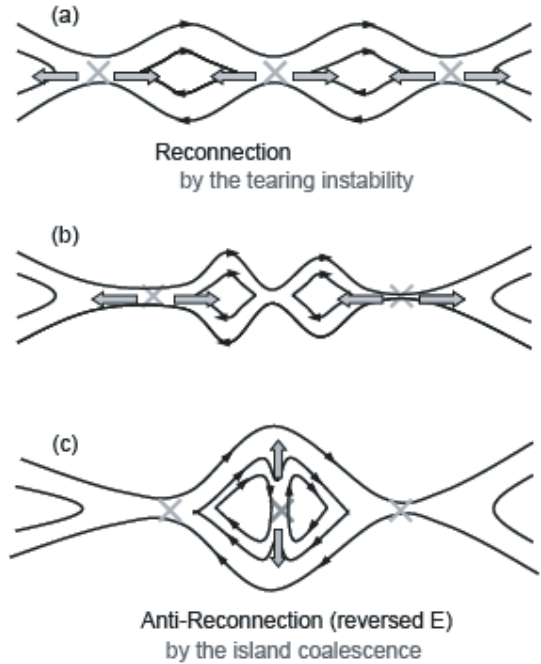}
\caption{Schematic illustration of multi-island coalescence. The thin cross marks indicate the X-lines of normal magnetic reconnection. The thick cross mark indicates the X-line of the anti-magnet-reconnection generated by the coalescence. The arrows indicate the flow directions.\label{fig:schematic}}
\end{figure}

Despite the intensive research of magnetic reconnection, it still remains unknown which of the two different type of acceleration is important for producing energetic electrons. In this respect, multi-island coalescence has drawn considerable attentions because it potentially contains both {\it X-Type} and {\it O-Type} mechanisms. A coalescence is the process of merging of two magnetic islands and has been studied by both theories and simulations \citep{finn77, pritchett79, biskamp80, tajima87, pritchett07, wan08}. Figure 1 shows a schematic illustration. In general, multiple number of X-lines and localized currents are generated from the tearing-mode instability (Figure \ref{fig:schematic}a). If we assume an X-line at the center was relatively weak compared to the other two X-lines, the two localized currents bounding the central X-line will eventually be attracted to each other by the Lorentz force (Figure \ref{fig:schematic}b). Each current is represented by magnetic islands and at the merging point of the two islands, a secondary magnetic reconnection or `anti-reconnection' \citep{pritchett08} occurs (Figure \ref{fig:schematic}c). The direction of the electric field of the anti-reconnection is reversed from the direction of the electric field of the primary reconnection. Finally, the two magnetic islands become one large magnetic island. 

While the first PIC simulation of particle acceleration during multi-island coalescence was performed more than decades ago \citep{tajima87}, a detailed study came out quite recently. 
\cite{pritchett08} showed that electrons are energized as the number of magnetic islands is reduced by coalescence. 
An important conclusion of the study is that the reversed electric field decelerates electrons near the anti-reconnection site. It was suggested that the main energization occurs through the {\it O-Type} mechanism. 

In this Paper, we extend the work by \cite{pritchett08} by performing 2D PIC simulations of multi-island coalescence with no guide field. The key and rather surprising finding of our simulations is that the anti-reconnection plays an important role in accelerating electrons. 
Very recently, \cite{tanaka10} also reported intense electron energization by the anti-reconnection, but our work  provides new insights from different perspectives on this issue because we fully analyzed the trajectories of accelerating electrons and clarified the importance of the anti-reconnection with respect to the other mechanisms.

The outline of the paper is as follows. In section 2, we describe the setup of the simulation runs and the overview of results. Section 3 presents analysis of electron energy spectra. Section 4 presents the trajectory analysis of energetic electrons and describe the details of their energization processes. In section 5, we summarize the results and discuss applications to the observations of the solar flares. Finally, section 6 contains the conclusion.

\section{Simulation Setup and Overview}

\begin{figure*}
\includegraphics[width=180mm]{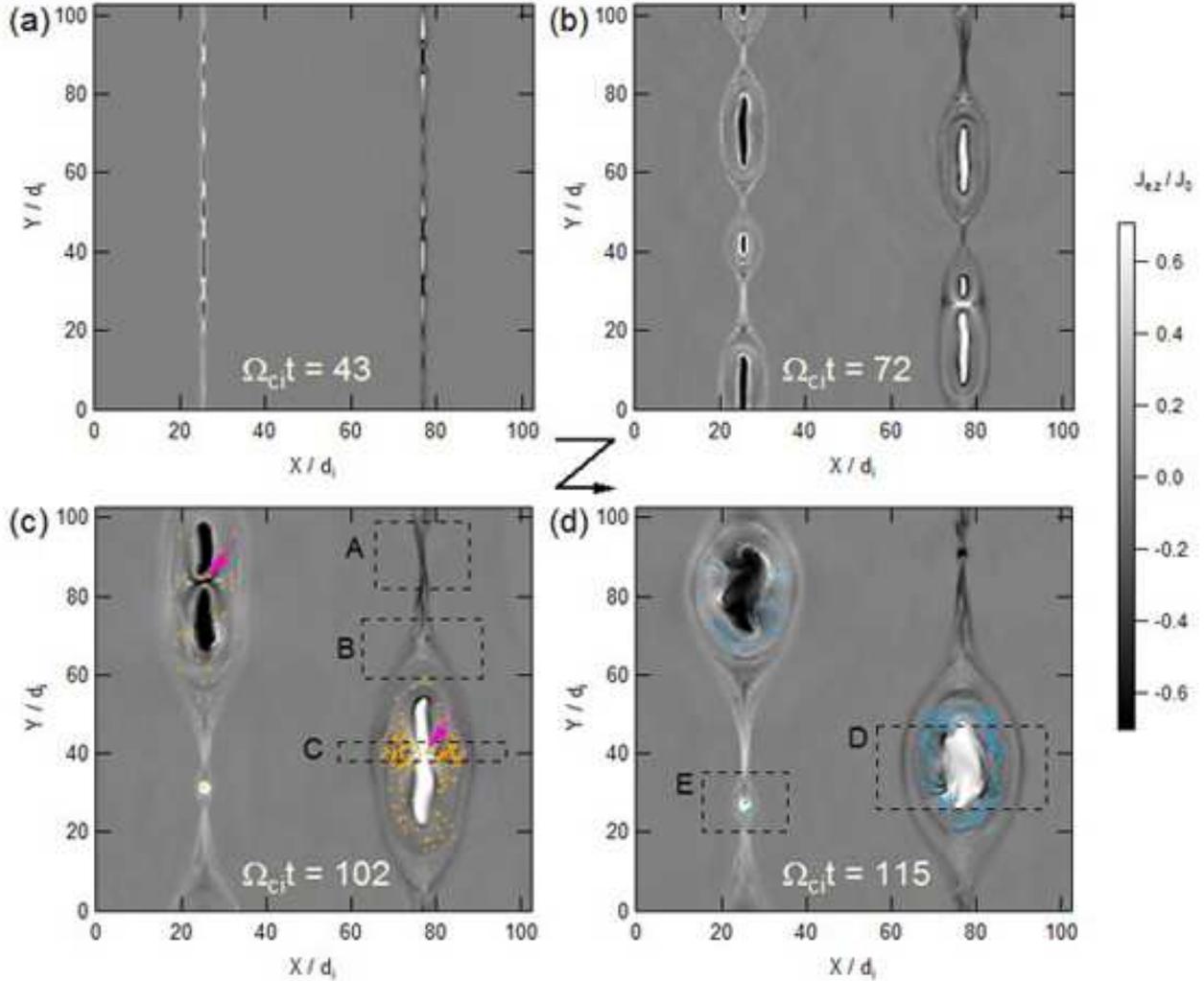}
\caption{Out-of-plane component of the electron current density obtained at (a) $\Omega_{\rm ci}$t = 42.7, (b) $\Omega_{\rm ci}$t = 70.0, (c) $\Omega_{\rm ci}$t = 102.0, (d) $\Omega_{\rm ci}$t = 114.7. The filled circles are the positions of the energetic electrons with energies (c) $\varepsilon\geq$1.2m$_{\rm e}c^2$ and (d) $\varepsilon\geq$1.4m$_{\rm e}c^2$. The arrows indicate island merging points. \label{fig:overview_csze}}
\end{figure*}
\begin{figure}
\epsscale{1.2}
\plotone{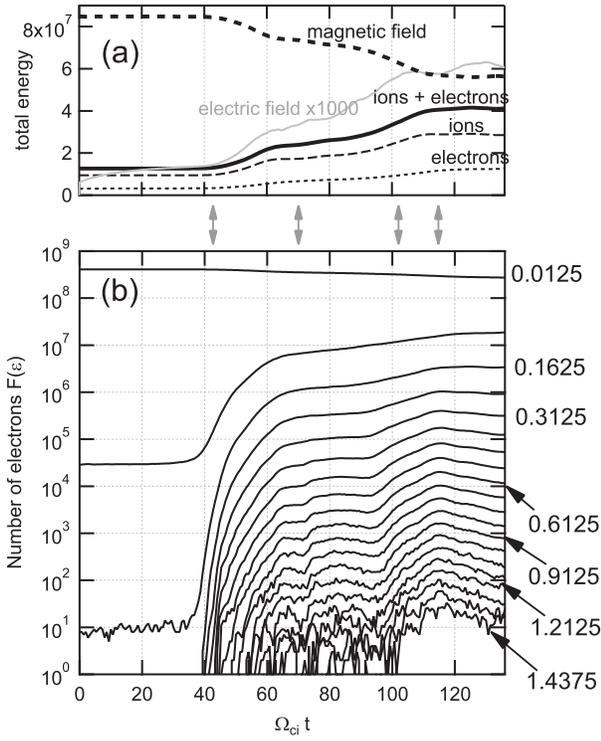}
\caption{(a) Time histories of particle and field energies integrated over the entire simulation domain. The electric field energy profile has been multiplied by 1000. (b) Stack plots of the time histories of the number of electrons counted in each energy bin $\varepsilon_i$ where $\varepsilon_i=$ (0.0125+0.025i)m$_{\rm e}c^2$, i=0,1,2..., from top to bottom. The bin size $\Delta\varepsilon_i$=$\pm$0.0125m$_{\rm e}c^2$. Some of the bin energies are annotated in the panel. The gray arrows in between the two panels indicate the time of the snapshots of Figure \ref{fig:overview_csze}.\label{fig:overview_timeprofile}}
\end{figure}
\begin{figure}
\epsscale{1.0}
\plotone{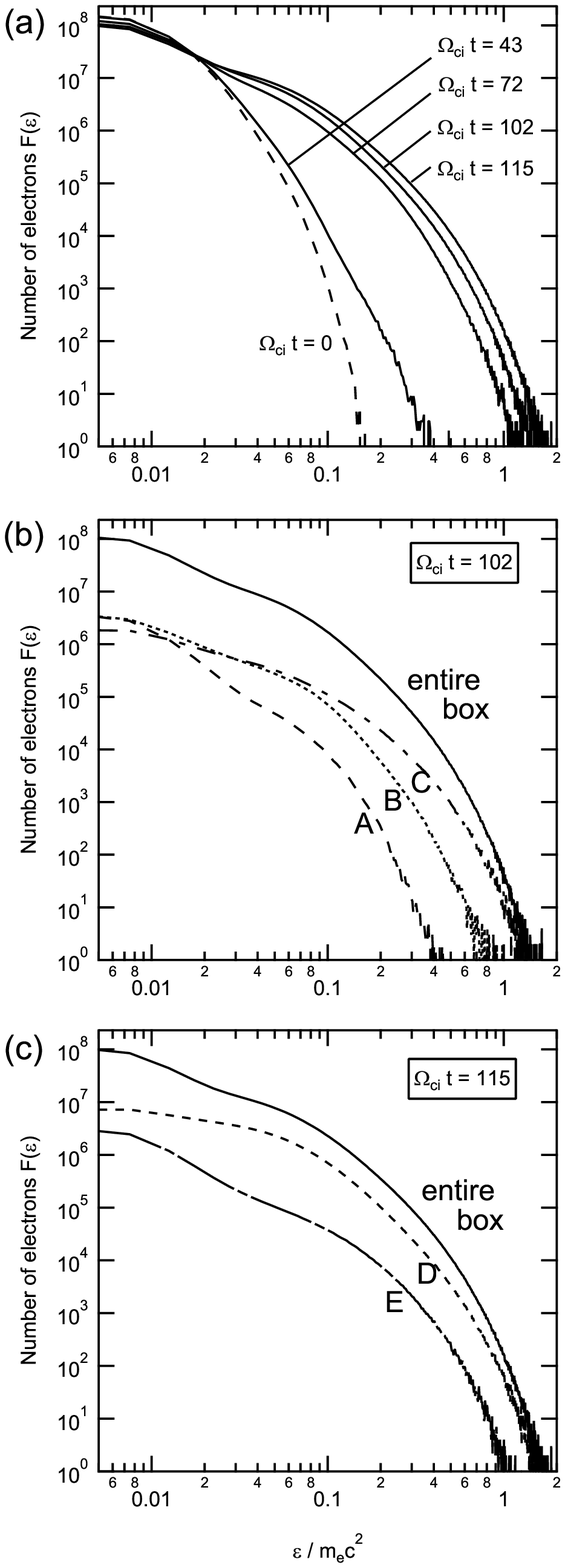}
\caption{(a) Time variation of the energy spectrum integrated over the entire simulation domain. (b) Energy spectra obtained at $\Omega_{\rm ci}$t=102 in the area A, B and C of Figure \ref{fig:overview_csze}c. (c)  Energy spectra obtained at $\Omega_{\rm ci}$t=115 in the area D and E of Figure \ref{fig:overview_csze}d. \label{fig:overview_pspe}}
\end{figure}

We utilize a two and half dimensional, fully relativistic PIC code \citep{hoshino87, shinohara01}. The initial condition consists of two Harris current sheets. The anti-parallel magnetic field is given by $B_y/B_0=\tanh((x-x_R)/D)-\tanh((x-x_L)/D)-1$ where B$_{\rm 0}$ is the magnetic field at the inflow boundary, $D$ is the half-thickness of the current sheet and $L_x$ and $L_y$ are the domain size in $\mathbf{\hat{x}}$ and $\mathbf{\hat{y}}$ direction, respectively. $x_L$=$L_x/4$ and $x_R$=3$L_x/4$ are the $x$-positions of the left and right current sheet, respectively.  Periodic boundaries are used in both directions. The ion inertial length $d_i$ is resolved by 25 cells. D=0.5$d_i$ and $L_x=L_y$=102.4$d_i$. The inflow, background plasma has the uniform density of $N_{B0}$=0.2$N_0$ where $N_0$ is the density at the current sheet center. The ion to electron temperature ratio is set to be $T_i/T_e$=5 for the current sheet and $T_i/T_e$=1 for the background. The background to current sheet temperature ratio $T_{bk}/T_{cs}$=0.1. The frequency ratio $\omega_{\rm pe}/\Omega_{\rm ce}$=3 where $\omega_{\rm pe}$ and $\Omega_{\rm ce}$ are the electron plasma frequency and the electron cyclotron frequency, respectively. The ion to electron mass ratio $m_i/m_e$=25 and the light speed $c$ is 15V$_A$ where V$_A$ is the Alfv\'{e}n speed defined as B$_0/\sqrt{4\pi N_0 m_{\rm i}}$. We used average of 64 particles in each cell. 297 particles per cell represents the unit density. No magnetic field perturbation is imposed at the beginning so that the system evolves from the tearing mode instability due to particle noise.

\begin{figure}
\epsscale{1.0}
\plotone{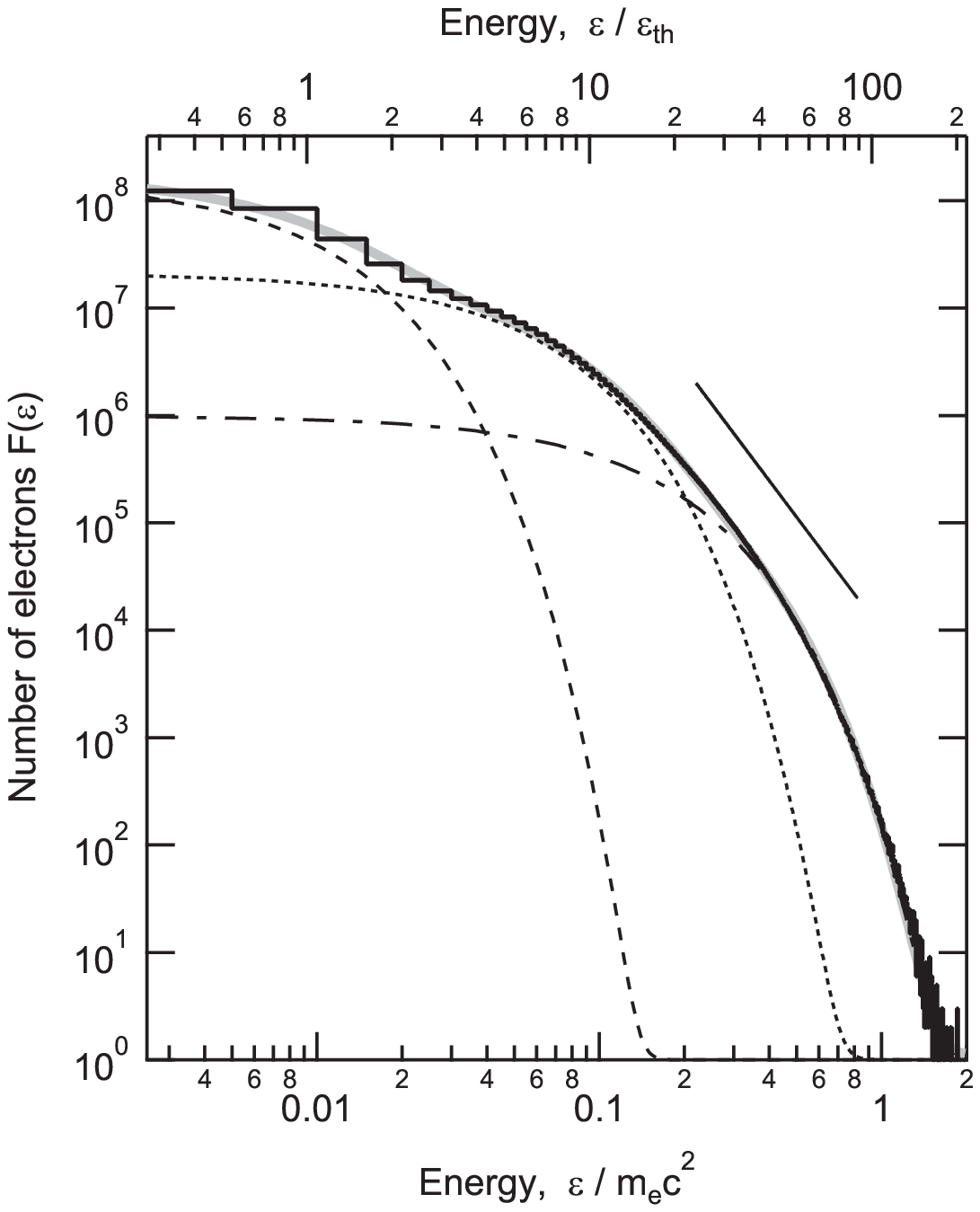}
\caption{Electron energy spectrum obtained at $\Omega_{\rm ci}$t=115. The gray curve shows the best fit model shown by the equation (\ref{eq:fit}). The dashed, dotted, and dash-dotted curves show the first, second, and third thermal component of the best fit model.  For the fitting procedure, we used the standard deviation of particle number count in each energy channel to evaluate the weight. The resultant numbers are F$_0$ = 1.12$\pm$0.15, A$_1$ = 1.52$\times$10$^{8}\pm$1.84$\times$10$^4$, T$_1$ = 7.3$\times$10$^{-3}$$\pm$1.3$\times$10$^{-6}$m$_{\rm e}$c$^2$, A$_2$ = 2.11$\times$10$^{7}\pm$6.96$\times$10$^3$, T$_2$ = 0.04$\pm$1.3$\times$10$^{-5}$m$_{\rm e}$c$^2$, A$_3$ = 9.90$\times$10$^{5}\pm$2.75$\times$10$^3$, T$_3$ = 0.11$\pm$8.4$\times$10$^{-5}$m$_{\rm e}$c$^2$, and $\chi^2$/d.o.f.=2.7$\times$10$^6$/350. For the horizontal axis, the energy $\varepsilon$ is normalized by the rest mass energy m$_{\rm e}$c$^2$ (bottom axis) and the initial background temperature $\varepsilon_{\rm th}$ (top axis). \label{fig:spectrum_obs}}
\end{figure}
%

The overall evolution of the system is very much the same as those of previous reports \citep{pritchett08}. The evolution of our simulation run is summarized in Figure \ref{fig:overview_csze} and \ref{fig:overview_timeprofile}. The linear growth of the tearing instability leads to the generation of at least seven X-lines by the time $\Omega_{\rm ci}$t=43 (Figure \ref{fig:overview_csze}a). In the non-linear stage, the number of X-line is reduced due to coalescence (Figure \ref{fig:overview_csze}b-d). At each merging point, anti-reconnection occurs and the out-of-plane component of the electron current density J$_{e,z}$ enhances at, for example, $(x,y)$=(25,84) and (77,40) in Figure \ref{fig:overview_csze}c. Small magnetic islands by the secondary tearing instability are also observed at, for example, $(x,y)$=(26,31) in Figure \ref{fig:overview_csze}c, but they will merge with a larger island. By the end of the simulation run, one large island remains in each current sheet.

Note that if two current sheets are sufficiently close, the tearing mode and magnetic reconnection drive each other \citep{pritchett80}. In the present case, the two current sheets evolve independently until $\Omega_{\rm ci}$t$\sim$75. In the later stages of the simulation ($\Omega_{\rm ci}$t$\gtrsim$105), the configuration of one layer becomes almost anti-symmetric to the other layer. However, by varying the separation between the two layers (or $L_x$), we verified that results presented in this paper are insensitive to the domain size.

Figure \ref{fig:overview_timeprofile}a shows the time profile of particle and field energies integrated over the entire simulation domain. Coalescence proceeds in two distinct stages: 40$\lesssim \Omega_{\rm ci}t \lesssim$65 and 90$\lesssim \Omega_{\rm ci}t \lesssim$115. During these intervals, the magnetic field energy is more rapidly converted to particle energy. At the end of the run, ions carry 67\% of the released magnetic field energy whereas electrons carry 33\%. About 34\% of the initial magnetic field energy has been released. Figure \ref{fig:overview_timeprofile}b shows the time profiles of the number of electrons of certain energies. It is apparent that higher energy flux increases rapidly in association with the two growth stages mentioned above. 
\begin{figure*}
\includegraphics[width=180mm]{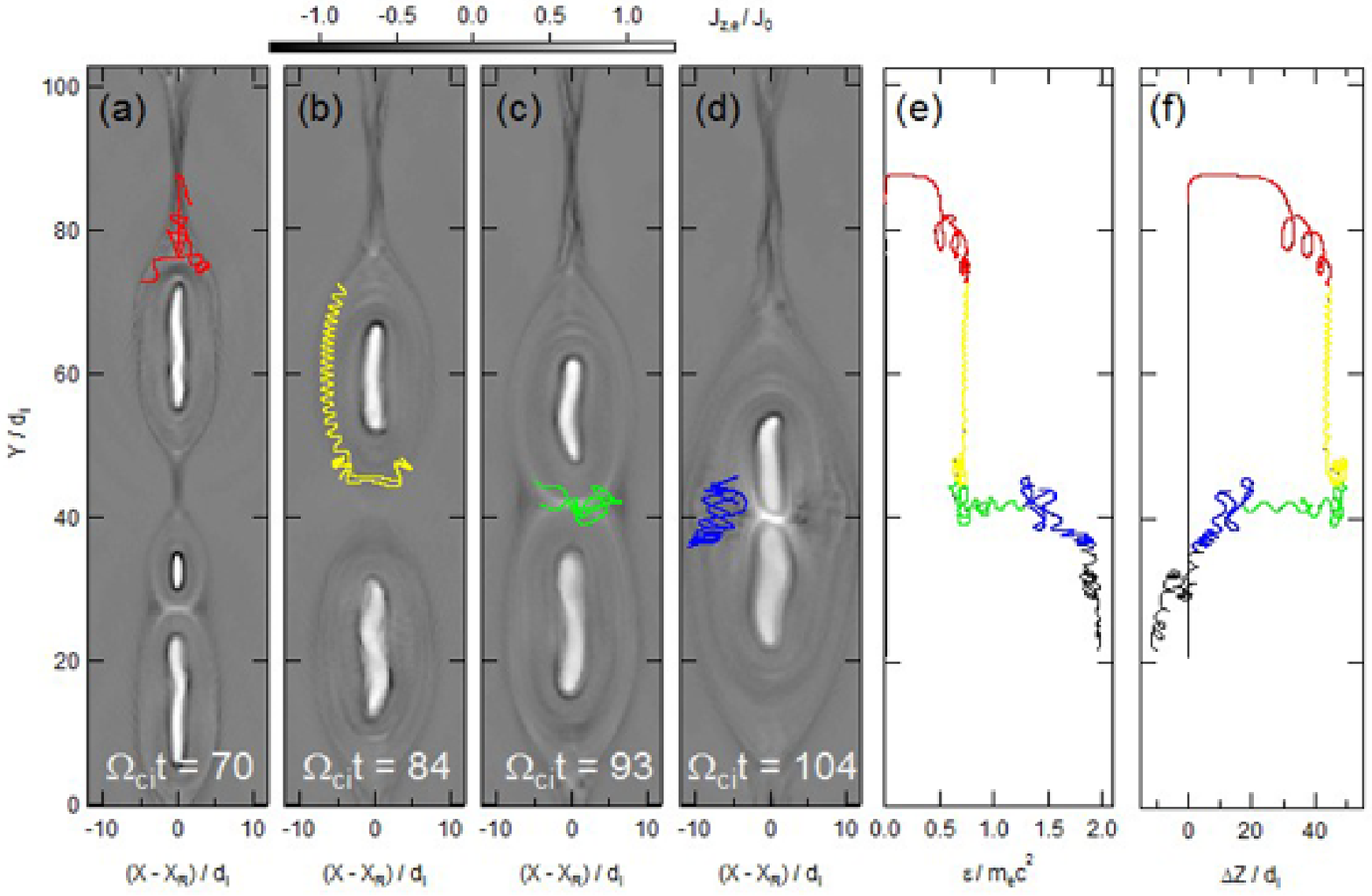}
\caption{\label{fig:XL-overview} The trajectory of particle \#1 during (a) 62.0$\leq\Omega_{\rm ci}t\leq$75.3, (b) 75.3$\leq\Omega_{\rm ci}t\leq$87.4, (c) 87.4$\leq\Omega_{\rm ci}t\leq$93.8, (d) 93.8$\leq\Omega_{\rm ci}t\leq$104.0. Also displayed are (e) the energy and (f) the displacement $\Delta$Z during 0$\leq\Omega_{\rm ci}t\leq$114.7 as functions of $Y$. The background images are the electron current density at (a) $\Omega_{\rm ci}$t=70.0, (b) $\Omega_{\rm ci}$t=84.0, (a) $\Omega_{\rm ci}$t=92.7, (a) $\Omega_{\rm ci}$t=104.0. }
\end{figure*}

In order to explore origins of the energetic electrons, we plotted locations of the most energetic electrons in Figure \ref{fig:overview_csze}c,d. Figure \ref{fig:overview_csze}c corresponds to the time of rapid increase of the energetic electron flux. At this time, electrons are distributed near the merging point suggesting that these electrons are created by the anti-reconnection. Figure \ref{fig:overview_csze}d corresponds to the time of the energetic electron peak flux. By this time, coalescence is almost complete and energetic electrons are uniformly distributed inside the merged island but surrounding the core region. 
It is also important to note that energetic electrons also exist in the secondary magnetic island. 

\section{Energy Spectrum}

Figure \ref{fig:overview_pspe}a shows the time variation of the electron energy spectrum integrated over the entire simulation domain. Electron heating starts during the earlier phase of the initial growth phase ($\Omega_{\rm ci}$t=43). By the end of the initial growth phase ($\Omega_{\rm ci}$t=72), the spectrum consists of at least two components: a cold thermal component representing the initial plasma condition and a hot thermal component representing the heated/accelerated plasma. This spectral form does not change very much there after but the heating/acceleration continues constantly ($\Omega_{\rm ci}$t=102, 115). Figure \ref{fig:overview_pspe}b shows the energy spectra integrated over the rectangular boxes in Figure \ref{fig:overview_csze}c. These are obtained when the highest energy electrons appeared around the anti-reconnection site ($\Omega_{\rm ci}$t=102). It is shown that the hot, energized electrons already exist in the vicinity of the X-line (A) but the magnetic field pile-up region produces more energetic electrons (B). The most energetic electrons, however, are created in the anti-reconnection region (C). Figure \ref{fig:overview_pspe}c shows the energy spectra integrated over the rectangular boxes in Figure \ref{fig:overview_csze}d. These are obtained when the flux of energetic electrons was at its peak ($\Omega_{\rm ci}$t=115). It is evident that the most energetic electrons exist in the merged island (D), although the secondary island alone can also produce a hot thermal component (E). The above results suggest that the most energetic electrons are first accelerated by the anti-reconnection and then further energized within the merged islands.

Figure \ref{fig:spectrum_obs} shows exactly the same spectrum shown in Figure \ref{fig:overview_pspe}a obtained at $\Omega_{\rm ci}$t=115, but it is now fitted by a best fit model:
\begin{eqnarray}
F(\varepsilon) &=& F_0 + A_1\exp{\left(-\frac{\varepsilon}{T_1}\right)} \nonumber \\
 &+& A_2\exp{\left(-\frac{\varepsilon}{T_2}\right)} + A_3\exp{\left(-\frac{\varepsilon}{T_3}\right)}
\label{eq:fit}
\end{eqnarray}
The first exponential component (T$_1\sim$7.3$\times$10$^{-3}$m$_{\rm e}$c$^2$, dashed curve) corresponds to the initial electron distributions. The second exponential component (T$_2\sim$4.2$\times$10$^{-3}$m$_{\rm e}$c$^2$, dotted curve) represents the heated electrons by the multi-island coalescence and associated magnetic reconnection. The third exponential component (T$_3\sim$0.11m$_{\rm e}$c$^2$, dash-dotted curve) may be viewed as either additional thermal component or a non-thermal component that was not able to extend to higher energy possibly because of the limited size of the simulation. 
For a reference, we indicated a slope of 3.5 by the solid line.

\section{Particle Trajectory}

Knowing that multi-island coalescence produces a very hot, possibly non-thermal component, we focus on the highest energy electrons. In order to clarify the exact process of electron energization, we traced backward in time the trajectories of the electrons with energies $\varepsilon>$1.4m$_{\rm e}$c$^2$ plotted in Figure \ref{fig:overview_csze}d, that is, in total, 198 electrons. 
Among these particles, we have found a variety of different energization processes and organized them by the type of acceleration mechanism.

\subsection{X-type Acceleration}

\begin{figure*}
\begin{center}
\includegraphics[width=170mm]{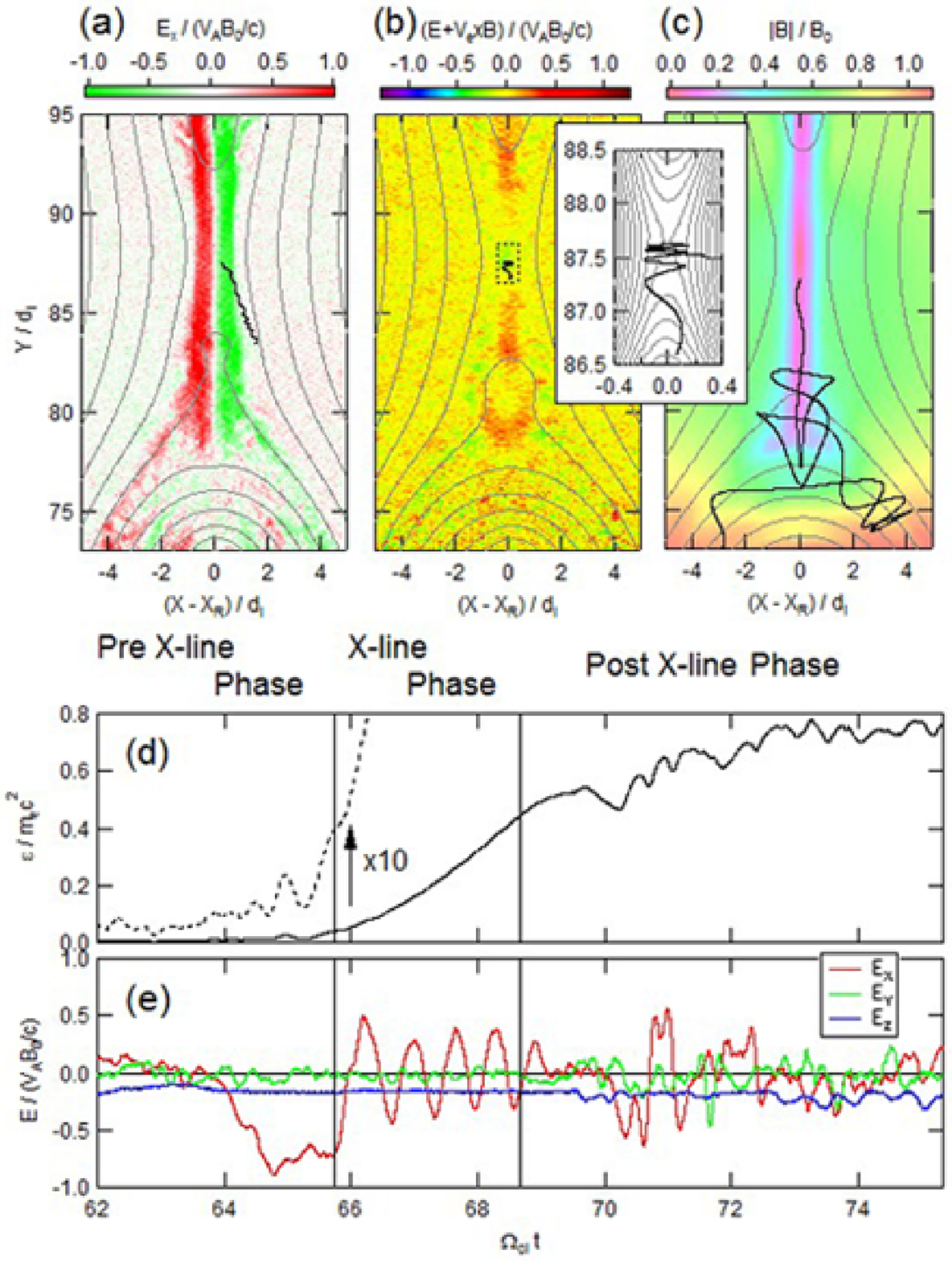}
\caption{\label{fig:XL-acc1} The enlarged view of the trajectory of particle \#1 (blue curves) during (a) 62.0$\leq\Omega_{\rm ci}t\leq$65.7 (Pre-X-line Phase), (b) 65.7$\leq\Omega_{\rm ci}t\leq$68.7 (X-line Phase) and (c) 68.7$\leq\Omega_{\rm ci}t\leq$75.3 (Post-X-line Phase), and the time histories of (d) the particle energy and (e) the electric field felt by the particle. The trajectory is superposed on images of (a) the out-of-plane component of the electric field, (b) the out-of-plane component of the non-motional electric field and (c) the magnetic field magnitude. The inset shows a blow-up of the X-line region of panel b. In each panel, the contour shows the magnetic field lines. The dashed curve in (d) is the same as the solid curve but multiplied by 10 to highlight the Pre-X-line Phase. In order to eliminate particle noise, the electric field profiles have been smoothed by a box average with the box size of $\Omega_{\rm ci}\Delta$t$\sim$0.13.}
\end{center}
\end{figure*}

Figure \ref{fig:XL-overview} shows the first example of the electron trajectories (hereafter Particle \#1). We divided the trajectory into four segments and plotted over the out-of-plane component of the electron current density. The whole profiles of energy and the displacement $\Delta$z are shown in the right two panels. The simulation itself is two-dimensional in $x$ and $y$ but the displacement $\Delta z$ can be calculated by integrating $v_z$ over time. The background images are just snapshots taken during each segment so care should be taken when comparing the trajectories with the images. The time of each image is chosen so that it best reflects the characteristics of each trajectory.

Particle  \#1 is first energized at one of the few X-lines formed within the first coalescence stage (Figure \ref{fig:XL-overview}a). Scatterings take place in the immediate downstream of the X-line, but the particle soon starts to travel along the rim of the island associated with the X-line and is further scattered when it reaches near the merging point of two islands approaching to each other (Figure \ref{fig:XL-overview}b). While wandering around the merging point, the coalescence progresses to form a region of localized current in which the particle is then confined to experience the second rapid energization (Figure \ref{fig:XL-overview}c). Note that this localized current corresponds to the anti-reconnection in which electric field direction is reversed. The energization persists even after the particle is ejected out of the central anti-reconnection region and the particle is pitch angle scattered afterwards (Figure \ref{fig:XL-overview}d). The first and second rapid energization occur in a very localized region ($\Delta y \sim$ 0.1 d$_i$) associated with the reconnection and anti-reconnection, respectively (Figure \ref{fig:XL-overview}e). What is important here is that the particle moves toward the positive $z$-direction up to $\Delta z \sim$50d$_i$, but by the reversed electric field, it turns its direction toward the negative $z$-direction. As a result, it passes through the original position and reaches  $\Delta z \sim$ -10d$_i$ by $\Omega_{\rm ci}$t=104.0.

Figure \ref{fig:XL-acc1} shows the details of the behavior of Particle \#1 during the first rapid energization. The trajectory is further divided into three time periods according to its characteristics. The time profiles of the particle energy as well as the electric fields felt by the particle are also displayed.

Figure \ref{fig:XL-acc1}a illustrates the period during which the particle approaches toward the X-line and starts to be energized. During this phase, we can identify the polarized electric field mainly parallel to the reconnection plane (i.e. $X-Y$ plane) and the particle is energized as it passes through this region. The polarized electric field felt by the particle reaches $\sim$ V$_{\rm A}$B$_{\rm 0}$ (Figure \ref{fig:XL-acc1}e). However, the corresponding energy increase is very small, $\Delta \varepsilon \sim$0.1m$_{\rm e}c^2$ (Figure \ref{fig:XL-acc1}d). In order to better visualize the energy increase, we multiplied the energy profile by 10, as shown by the dashed curve.

Figure \ref{fig:XL-acc1}b illustrates the `X-line Phase' during which the particle stays at the very center of the diffusion region or the X-line and is rapidly energized. Within a relatively short period of time $\Omega_{\rm ci}\Delta$t$\sim$2, the energy increases nearly an order of magnitude. Note also that this energization takes place in the so-called inner diffusion region. Recent simulations revealed that the out-of-plane current is localized only at the center of the diffusion region (`inner diffusion region') so that the reconnection rate remains fast while electrons form a  high-velocity jet in the `outer diffusion region' \citep{daughton06, fujimoto06, shay07, karimabadi07, phan07}. Electrons are unmagnetized in these regions so that a snapshot of the $z$-component of E+V$_e\times B$ makes clear in which region the electron can be energized. The inner diffusion region (colored blue) is at the X-line and the outer diffusion region (colored red) extends toward the downstream.  It is evident that the electron is rapidly energized within the `inner diffusion region'. A close look at the particle trajectory reveals a meandering motion in this region (Figure \ref{fig:XL-acc1} inset). 

Figure \ref{fig:XL-acc1}c illustrates the `Post-X-line Phase' during which the particle moves toward the downstream and is pitch angle scattered. The rate of energy gain decrease, but the particle is energized up to $\sim$0.7m$_{\rm e}c^2$. The energization takes place near the edge of the outer diffusion region or the outflow exhaust. This is where magnetic field magnitude starts to increase. Also, because electrons are already pre-energized at the X-line, they have gyro-radii comparable to the curvature of the magnetic field lines of this region. Concurrently, the gradient B and curvature B drift motion plays an important role for the additional energization during this phase. In the later half of the Post-X-line Phase, Particle \#1 is reflected by a mirror force at ($x,y$)$\sim$(3.5d$_i$, 74d$_i$). The particle again passes through the edge of the diffusion region at ($x,y$)$\sim$(0, 76d$_i$) but the energy increase is very small. Finally, it escapes away from the diffusion region.

\begin{figure*}
\includegraphics[width=170mm]{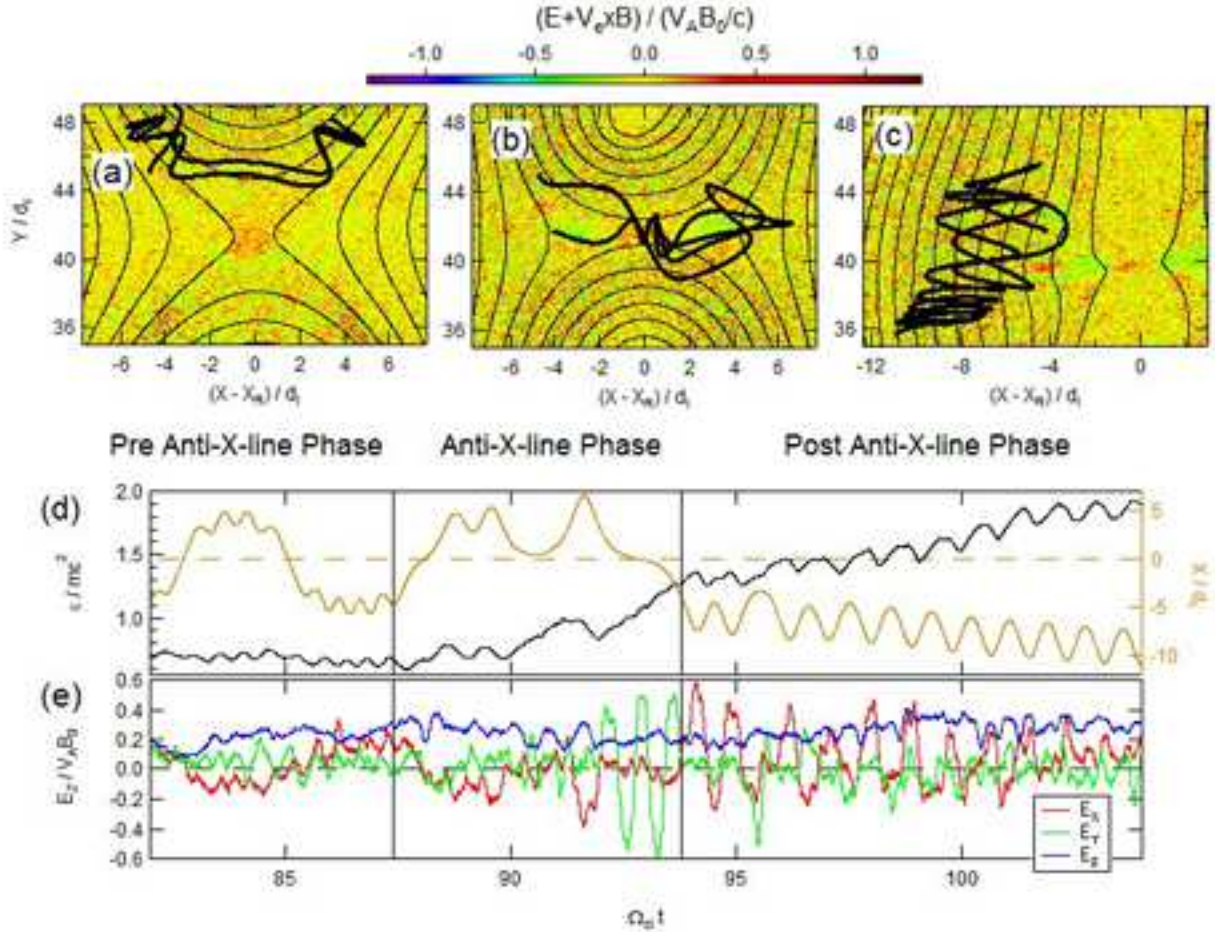}
\caption{\label{fig:XL-acc2} The enlarged view of the trajectory of particle \#1 during (a) 75.3$\leq\Omega_{\rm ci}t\leq$87.4 (Pre-Anti-X-line Phase), (b) 87.4$\leq\Omega_{\rm ci}t\leq$93.8 (Anti-X-line Phase) and (c) 93.8$\leq\Omega_{\rm ci}t\leq$104.0 (Post-Anti-X-line Phase), and the time histories of (d) the particle energy, $x$ positions and (e) the electric field felt by the particle. The trajectories in (a-c) are superposed on images of the out-of-plane component of the non-motional electric field with the contours showing the magnetic field lines.  In order to eliminate particle noise, the electric field profiles have been smoothed by a box average with the box size of $\Omega_{\rm ci}\Delta$t$\sim$0.13.}
\end{figure*}

We now explore the details of the second rapid energization of Particle \#1. Figure \ref{fig:XL-acc2} shows the last three of the four phases described in Figure \ref{fig:XL-overview}. The enlarged views of the trajectory are shown in the upper three panels and the time profile of the particle energy as well as the electric field felt by the particle are shown in the lower two panels.

Figure \ref{fig:XL-acc2}a illustrates the `Pre-Anti-X-line Phase' during which the particle approaches toward the merging point of two large islands. The particle is reflected twice by the mirror force but is not energized. The reversed electric field is already large at this phase reaching to $\sim$0.3V$_{\rm A}$B$_{\rm 0}$ and the non-magnetized region starts to appear at the merging point.

Figure \ref{fig:XL-acc2}b illustrates when the anti-reconnection is taking place (`Anti-X-line Phase'). The particle is drawn into the anti-reconnection region and is energized significantly. The duration of energization, however, is somewhat longer than the time duration of the first rapid energization ($\Omega_{\rm ci}\Delta$t$\sim$6). Note that the anti-reconnection is embedded in closed magnetic field lines and the outflow from the anti-X-line collides with these magnetic fields. Moreover, because of the first rapid energization, the gyro-radii of the electron is large comparable to the scale of the anti-reconnection. The electron motion is thus decoupled from the magnetic field lines. The above features lead to trapping of the electron within the anti-reconnection region and hence significant energy gain. Yet, each `kick' occurs in the inner diffusion region as was the case during the first rapid energization. This is more clearly shown in the time profiles of the particle energy and the $x$ positions (Figure \ref{fig:XL-acc2}d). The oscillating feature with the time scale of $\Omega_{\rm ci}\Delta$t$\sim$1 corresponds to the gyro-motion of the particle. During the Anti-X-line Phase, this oscillatory feature is modified and a continuous energy increase occurs whenever the $x$-position is within $\pm$1d$_i$ from the X-line.

Figure \ref{fig:XL-acc2}c shows the `Post-Anti-X-line Phase'. Note the change of the horizontal axis range. The particle is ejected out of the anti-reconnection region but continues to gain energy by the gradient B and curvature B drift acceleration. Because of the contracting motion of the merged island, the electric field in this region continues to increase and reaches 0.4V$_{\rm A}$B$_{\rm 0}$. As a result, the particle energy reaches near 2m$_{\rm e}$c$^2$.

In summary, Particle \#1 is first energized at and around the X-line generated during the non-linear evolution of the tearing mode instability. It is then energized at and around the anti-X-line generated by the merging of two islands. In both cases, the particle is accelerated directly by the reconnection electric field. As a result, the energy reaches more than $\varepsilon>$m$_{\rm e}c^2$. 

\subsection{O-type Acceleration}

\begin{figure*}
\includegraphics[width=170mm]{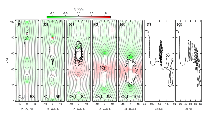}
\caption{\label{fig:OL-Drake1} The trajectory of particle \#2 during (a) 46.7$\leq\Omega_{\rm ci}t\leq$56.0, (b) 56.0$\leq\Omega_{\rm ci}t\leq$62.0, (c) 62.0$\leq\Omega_{\rm ci}t\leq$97.3, (d) 97.3$\leq\Omega_{\rm ci}t\leq$99.3 and (e) 99.3$\leq\Omega_{\rm ci}$t$\leq$114.7. Also displayed are (f) the energy and (g) the displacement $\Delta$Z during 0$\leq\Omega_{\rm ci}t\leq$114.7 as functions of $Y$. The background images are the out-of-plane component of the electric field at (a) $\Omega_{\rm ci}$t=52.7, (b) $\Omega_{\rm ci}$t=58.7, (c) $\Omega_{\rm ci}$t=84.0, (d) $\Omega_{\rm ci}$t=98.0 and (e) $\Omega_{\rm ci}$t=104.0.}
\end{figure*}

We now show another example of electrons that behaved quite differently from Particle \#1 (hereafter Particle \#2). Figure \ref{fig:OL-Drake1} shows the trajectory with the same format as Figure \ref{fig:XL-overview} but the trajectory is divided into five segments instead of four and the background images are replaced by the out-of-plane component of the electric field instead of the electron current density. 

The first energization occurs at $y\sim$88.5d$_i$ where an X-line develops as a consequence of the first coalescence growth phase (Figure \ref{fig:OL-Drake1}a). There is a small, secondary island at $y\sim$86d$_i$ but it only modulates the particle orbit and plays a minor role for the energization. The particle travels until it reaches another X-line which is also produced by the first coalescence growth (Figure \ref{fig:OL-Drake1}b). There, the particle receives two `kicks' by the reconnection electric field. The second kick occurs because the particle is reflected by the magnetic fields being piled-up on the pre-existing current sheet and comes back to where it can again receives energy from the reconnection electric field. The two kicks can be identified in Figure \ref{fig:OL-Drake1}f at ($\varepsilon/$m$_{\rm e}$c$^2$, y/d$_{\rm i}$)$\sim$(0.5, 45) and $\sim$(1.8, 50). 

What makes Particle \#2 different from \#1 is the behaviour during the rest of the trajectory. After the two kicks, the particle is trapped within an island and circulates 6 times (Figure \ref{fig:OL-Drake1}c), as can be counted from the zigzag orbit of the $\Delta z$-profile (Figure \ref{fig:OL-Drake1}g). Note the fact that this island is moving toward the negative $y$-direction because this island and the other island at $y\sim$20d$_i$ are attracting to each other. Thus, the direction of the gradient B and curvature B drift is positive $z$ at both end of the island while the direction of the motional electric field is negative and positive at the upper ($y\sim$70d$_i$) and lower ($y\sim$50d$_i$) ends, respectively. As a result, the particle gains energy at the upper end but loses energy at the lower end. After all, the net energy gain of the particle is kept small. This is represented by the foldings and overlappings of the curve in Figure \ref{fig:OL-Drake1}f at $\varepsilon\sim$1m$_{\rm e}c^2$.

The particle eventually reaches the merging point where it receives energy at the anti-X-line (Figure \ref{fig:OL-Drake1}d), but the anti-X-line has already been well developed so that the particle is not trapped in this region. The particle soon exits from the merging region and start to circulate inside the newly created, large island (Figure \ref{fig:OL-Drake1}e). This large island is contracting in this phase so that the particle can gain energy every time it passes through the both end of the island \citep{drake06}. There exists reversed electric field in between the island edges so that the particle can gain energy through the gradient B and curvature B drift.   As for the displacement along the $z$-direction, the particle moves toward negative $z$-direction only during the period of energization by the anti-reconnection. The resultant displacement is $\Delta z\sim$100d$_i$.

In summary, Particle \#2 gains significant amount of energy from the motional electric field due to the contracting motion of the merged island. The final energy reaches as high as that of Particle \#1, that is $\varepsilon\sim$2m$_{\rm e}c^2$.

Another important feature of coalescence is the bouncing motion of merged islands \citep{tajima87, wan08c}. The merged island still possesses a certain amount of kinetic energy so that a contracting motion of an island is followed by an expanding motion. 

\begin{figure*}
\includegraphics[width=170mm]{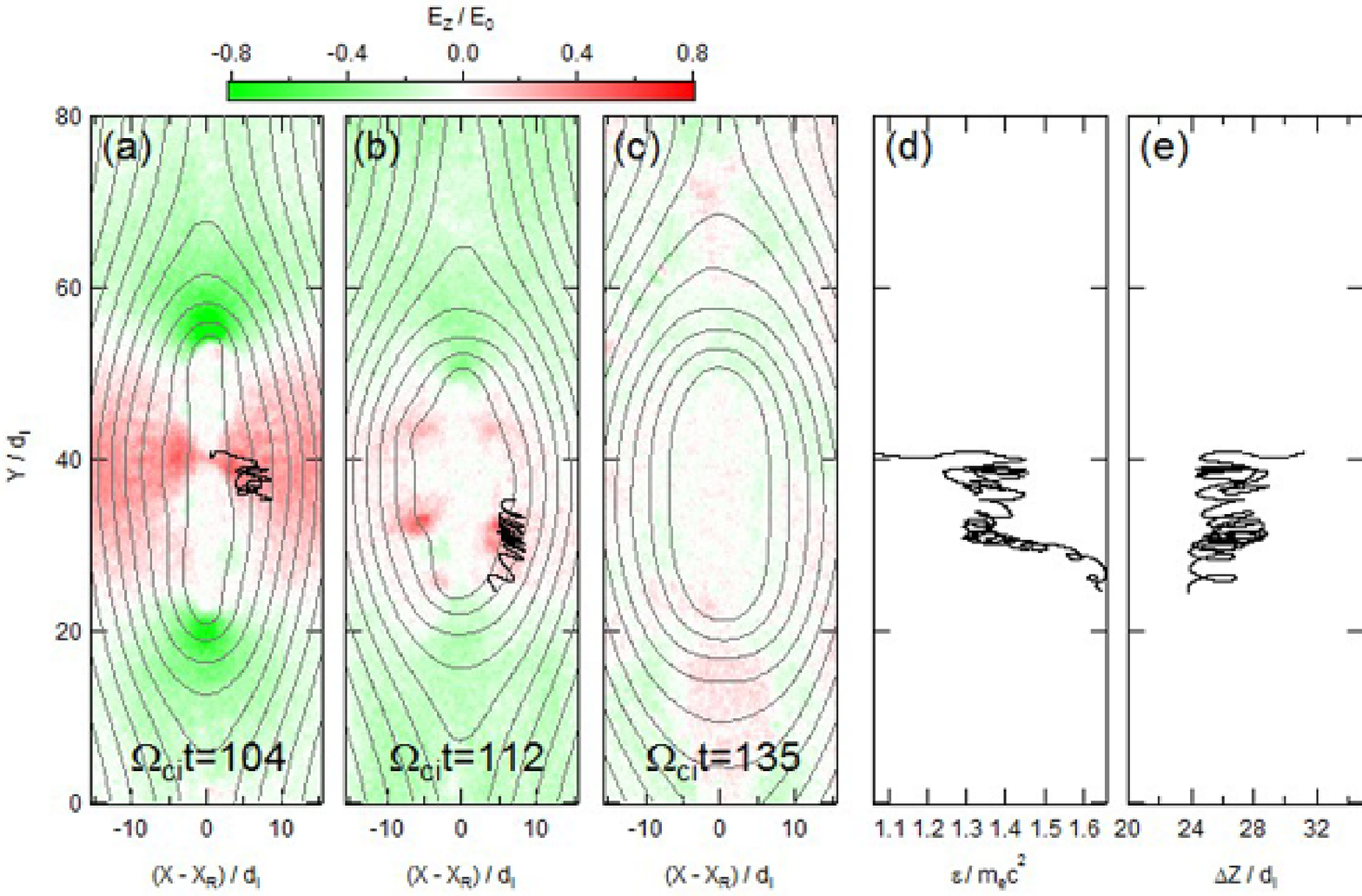}
\caption{The trajectory of particle \#3 during (a) 100.3$\leq \Omega_{\rm ci}$t$\leq$106.0 and (b) 106.0$\leq \Omega_{\rm ci}$t$\leq$114.7. The energy (d) and the displacement (e) during these periods are shown in the same panel. The trajectory after $\Omega_{\rm ci}$t=114.7 is not shown. The background images are the out-of-plane component of the electric field at when the merged island is (a) contracting, $\Omega_{\rm ci}$t=104.0, (b) stopped contracting, $\Omega_{\rm ci}$t=112.0, and (c) expanding, $\Omega_{\rm ci}$t=135.6. The contour shows the magnetic field lines.
\label{fig:OL-Drake2}}
\end{figure*}

Figure \ref{fig:OL-Drake2}(a-c) shows the evolution of the merged island. At $\Omega_{\rm ci}$t=104, the anti-reconnection has stopped but the reversed E$_z$ electric field remains because of the expanding motion in the $x$-direction (Figure \ref{fig:OL-Drake2}a). The shape of the merged island is somewhat elongated in the $y$-direction and the contracting motion in the $y$-direction still continues as evident from the negative E$_z$ region.  At $\Omega_{\rm ci}$t=112, the merged island has already expanded in the $x$-direction and the contracting motion in the $y$-direction has become weak (Figure \ref{fig:OL-Drake2}b). Notable here is the ripple structures that surround the island core region, as evidently shown by the enhanced, localized E$_z$ regions. These structures are due to a turbulent motion within the island. The turbulent motion is generated because the converging flows from the top and the bottom are mixing with each other.  The ripple structures propagate mainly in the $y$-direction away from the merging center. By the time $\Omega_{\rm ci}$t=135, the ripple structure disappears and the island has become round. It is very slowly expanding. 

Also shown in Figure \ref{fig:OL-Drake2} is a sample trajectory of electrons that pass through a  ripple structure (hereafter Particle \#3). After being energized up to $\varepsilon\sim$1.3m$_{\rm e}c^2$, Particle \#3 undergoes pitch angle scattering just outside an anti-X-line region, i.e. ($x$,$y$)$\sim$(6,38) (Figure \ref{fig:OL-Drake2}a). It then travels toward the negative $y$ direction (Figure \ref{fig:OL-Drake2}b). During this travel, the particle encounters the ripple structure identified at ($x$,$y$)$\sim$(6,31). This electric field is generated by a local flow toward the positive $x$ direction and the local magnetic field directed toward the negative $y$-direction. The electron is energized by the enhanced E$_z$ in association with the gradient B / curvature B drift (Figure \ref{fig:OL-Drake2}d). A gradient E drift is unlikely to be causing the energization because such drift motion is perpendicular to E$_z$, the only dominant component at this time.


\subsection{Hybrid Type Acceleration}

In a separate paper, we reported an `island surfing' mechanism of electron acceleration during magnetic reconnection \citep[][]{oka10b}. This mechanism utilizes the secondary magnetic islands that are produced in the diffusion region.  Inside each island, electrons are trapped for a significant period of time so that they are energized continuously by the reconnection electric field prevalent in the diffusion region. Although a pre-acceleration is required to keep electrons trapped within islands, the trapped electrons can receive unlimited amount of energy as long as the island stays in the diffusion region. 

Note again that while {\it X-Type} acceleration takes place at any X-line, {\it O-Type} acceleration occurs far away from X-lines by taking advantage of a closed field line geometry. The `island surfing' mechanism takes place in a diffusion region very close to the X-line, but at the same time, makes use of the closed field line geometry of a secondary magnetic island. As such, this mechanism falls into both categories described above and we rather consider it as an hybrid type. 

The `island surfing' is a natural consequence of the secondary tearing instability of magnetic reconnection, and hence not directly related to multi-island coalescence. Therefore, we do not discuss its detail any further in this paper. 


\section{Discussion}

\subsection{Acceleration processes}

We performed 2D PIC simulations of multi-island coalescence with no guide field. By following the trajectories of 198 energetic electrons, we identified various energization mechanisms as have been summarized in Table \ref{tbl}. Note again that these electrons are extracted at the peak time of energetic electron flux with the criteria of $\varepsilon>$1.4m$_{\rm e}$c$^2$. The first column shows the categorization of each mechanism. The second column shows mechanism nomenclature used throughout this Paper. The third column shows the origin of the electric field from which particles gain their energies. The fourth column shows what we refer to as the `contribution probability' P$_c$ defined as $P_c = 100n_{acc}/n_{tot}$ where $n_{acc}$ is the number of electrons accelerated by the mechanism and $n_{tot}$ is the total number of electrons we analyzed.  In the present case, $n_{tot}$=198. For each trajectory of the 198 electrons, we checked which mechanism the particle had been energized through and summed up the number of trajectories each mechanism worked on. Because each particle experiences more than one different mechanism, the total of P$_c$ does not equal to 100\%. The fifth column shows the largest energy gain by each mechanism $\Delta\varepsilon_l$. Finally, the last column of Table \ref{tbl} indicates which literature discussed each mechanism.  Below, we summarize each mechanism based on our simulation results and describe how we counted $n_{acc}$ for each mechanism. 

The `surfing' was difficult to identify in our simulation because of the negligible amount of energy increase ($\Delta\varepsilon_l$/m$_{\rm e}$c$^2<$0.1). This may be due to the undriven nature of our simulation that generates polarization electric field  E$_{\rm p}\sim$1V$_AB_0$. The original study of the `surfing' mechanism used a driven magnetic reconnection that generates the polarization electric field of E$_{\rm p}\sim$6V$_AB_0$ so that the Lorentz force can be balanced by the force from E$_{\rm p}$.
For Table \ref{tbl}, we simply did not count the number of the `surfing' mechanism.

The `X-line' mechanism is the classical way of energizing electrons, but based on the recently revealed two-scale structure, we verified that a significant energization takes place at the very center of the diffusion region or the `inner diffusion region'  ($\Delta\varepsilon_l$/m$_{\rm e}$c$^2\sim$0.8). This may be a matter of course given the fact that the out-of-plane current is localized within the inner diffusion region. The number of electrons energized by this mechanism $n_{acc}$ was 117. When counting this mechanism, we made sure that the electrons passed through the inner diffusion region.

The `Anti-X-line' mechanism is the most common energization mechanism among the trajectories we analyzed ($n_{acc}$=169). It is often intense and and the largest energy increment was $\Delta\varepsilon_l$/m$_{\rm e}$c$^2\sim$1.8. The energization process itself is basically the same as the `X-line' mechanism but there are three major differences. One is that the anti-X-line is created in a driven manner by two separate magnetic islands approaching toward each other. Since each island is expelled in association with the outflow from an X-line, the anti-reconnection rate reaches as large as 1 V$_{\rm A}$B$_{\rm 0}$. The second major difference is that an anti-X-line is bounded by the closed field lines of the merged island. The size of the anti-reconnection region is relatively small but the electrons can easily be trapped at and around the anti-reconnection region. This effect leads to multiple number of interaction with the inner diffusion region of the anti-reconnection. As a result, the `Anti-X-line' mechanism was found to be a significant energization mechanism, although the anti-reconnection is a transient process that takes place in a small region. The third major difference between the X-line and the. anti-X-line is that the electric field is reversed at the anti-X-line. Because of this, electrons move toward the opposite direction while being energized. In fact, this feature was what we used to count the number of anti-X-line mechanism in Table \ref{tbl}.

The `grad-B/curv-B drift' mechanism is also a common process for electron energization ($n_{acc}\sim$174) but each kick is not significant ($\Delta\varepsilon$/m$_{\rm e}$c$^2<$0.3). Typically, this process occurs after a particle exits from either X-line or anti-X-line. We identified this mechanism whenever the trajectory showed energy increase during a drift motion near either X-line or anti-X-line. A drift motion can also take place at each end of magnetic island but, in such cases, the trajectory can be regarded as a part of circulation inside the island and are not counted as the `grad-B/curv-B drift' mechanism. The number of the `grad-B/curv-B drift' mechanism $n_{acc}$ was 174.

The `contracting island' mechanism was identified in 79 trajectories. Since this mechanism requires energy gain at both end of island, the process appears only in the later phase of island coalescence. In the earlier phase of coalescence, the merging is not complete and either of the two approaching island is not contracting. We identified this mechanism by the zigzag orbit of the $y$-$\varepsilon$ plot. The largest energy gain was $\Delta\varepsilon_l$/m$_{\rm e}$c$^2\sim$0.6.

\begin{deluxetable*}{lllccr}
\tabletypesize{\scriptsize}
\tablecaption{Electron acceleration mechanisms during multi-island coalescence with no guide field. \label{tbl}}
\tablewidth{0pt}
\tablehead{
\colhead{ } & \colhead{Mechanism} & \colhead{Electric field origin} & \colhead{P$_{\rm c}, $\%} & \colhead{ $\Delta\varepsilon_{l}$/m$_{\rm e}$c$^2$} &
\colhead{Reference}
}
\startdata
X-Type &surfing             &reconnection      &N/A& $<$0.1  &\cite{hoshino05} \\
       &X-line              &reconnection       &59&0.8& \cite{hoshino01} and references therein \\
       &anti-X-line         &anti-reconnection  &85&1.8& Oka et al. (This paper); \cite{tanaka10}\\
       &grad-B/curv-B drift &(anti-)reconnection&88&0.3& \cite{hoshino01}  \\
O-Type &contracting island  &island contraction &40&0.6& \cite{drake06} \\
       &ripple              &island ripples     &31&0.3& Oka et al. (This paper) \\
hybrid &island surfing      &reconnection       & 2&0.6& \cite{oka10b}
\enddata
\end{deluxetable*}

The `ripple' mechanism was found to be as important as the `contracting island' mechanism, $n_{acc}$=61. During the bouncing motion of a newly merged island, a turbulent flows appear inside the island so that many electrons pass through localized electric fields. We counted the number of this mechanism by checking the association of an energy increase with the localized electric field within a merged island.

The `island surfing' mechanism was very rare. Only 4 electrons were found to be energized by this mechanism. This is probably because of the relatively small reconnection electric field in the later phase of the simulation run when secondary islands are created. Moreover, the size of each X-line was limited due to the presence of magnetic islands. If an X-line were able to reach a steady state, it would continue to spawn a number of secondary islands, which situation would lead to more importance of this mechanism.

In Table \ref{tbl}, we classified each mechanism as either {\it X-Type}, {\it O-Type}, or a hybrid. An energization mechanism belongs to the {\it X-Type} if it occurs in the diffusion region of magnetic reconnection and/or in the magnetic field pile-up region just downstream of the diffusion region. The source of energy is the reconnection electric field. On the other hand, an energization mechanism belongs to the {\it O-Type} if it takes place within the closed geometry of magnetic field or magnetic islands. The source of energy is the kinetic motion of the island. A hybrid mechanism refers to an energization within a secondary island located in the diffusion region.

\subsection{Displacement along the out-of-plane direction}

The striking feature of electron acceleration at the anti-reconnection site is the direction of motion in the $z$-direction being opposite to that of electrons at the primary reconnection site. If electrons were magnetized and drifting, they lose their energy by the reversed electric field \citep{pritchett08}. However, at the very center of the anti-reconnection, electrons are not magnetized and not drifting so that they can move parallel to the reversed electric field and receive substantial amount of energy.

\begin{figure}
\epsscale{1.2}
\plotone{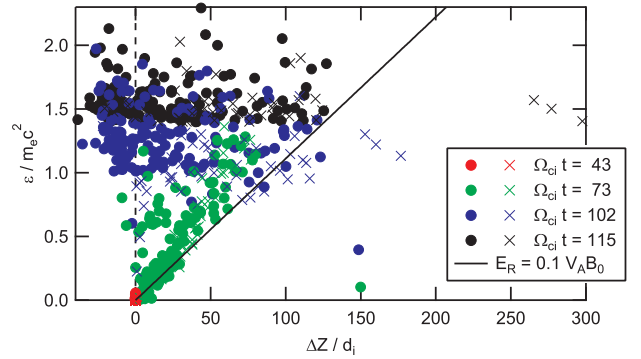}
\caption{The displacement of the most energetic electrons along the $z$-direction at different times. The filled circles and the crosses are for the electrons in the right and left current sheet, respectively. The sign of $\Delta z$ is converted so that electrons of both right and left current sheet move toward the same direction as that of the primary reconnection electric field. \label{fig:diffusion}}
\end{figure}

Figure \ref{fig:diffusion} shows the displacement of the most energetic electrons along the out-of-plane direction, $\Delta z$, obtained at $\Omega_{\rm ci}$t = 43, 72, 102, and 115. The solid line indicates a hypothetical energization by the electric field of 0.1V$_{\rm A}$B$_{\rm 0}$ which roughly corresponds to the primary reconnection electric field measured in the earlier phase of the simulation. It is also the typical value of steady state magnetic reconnection \citep{shay07}. In the earlier phase of the simulation run ($\Omega_{\rm ci}$t=73, green marks), many electrons follow this solid line because they are energized at the primary reconnection regions. They can reach as far as 100 d$_{\rm i}$. Later on, electrons are accelerated by the anti-reconnection so that they move toward the opposite direction ($\Omega_{\rm ci}$t=102, blue marks). By the time of peak electron flux ($\Omega_{\rm ci}$t=115, black marks), the asymmetry of the particle distribution remains,  but most particles are confined within 140 d$_{\rm i}$ from the initial positions.

The fact that many electron positions are largely deviated from the line of energization by the primary reconnection, particularly in the later phase of the simulation run, indicates that these electrons experienced anti-reconnection.  Some electrons showed unusually large displacement of $\Delta z>$140d$_{\rm i}$. All of these particles are energized by the `island surfing' mechanism and so did not experience anti-reconnection and direction turnings. 

\subsection{Application to the Solar Flares}

Let us now discuss applications of our simulations. We used periodic boundary conditions under the assumption of infinitely long current sheet in the $y$-direction. Therefore, our simulation can probably be best applied to the solar flares which show evidence of very long current sheets \citep[e.g.][]{sui03, bemporad06}.

\begin{figure*}
\begin{center}
\includegraphics[width=160mm]{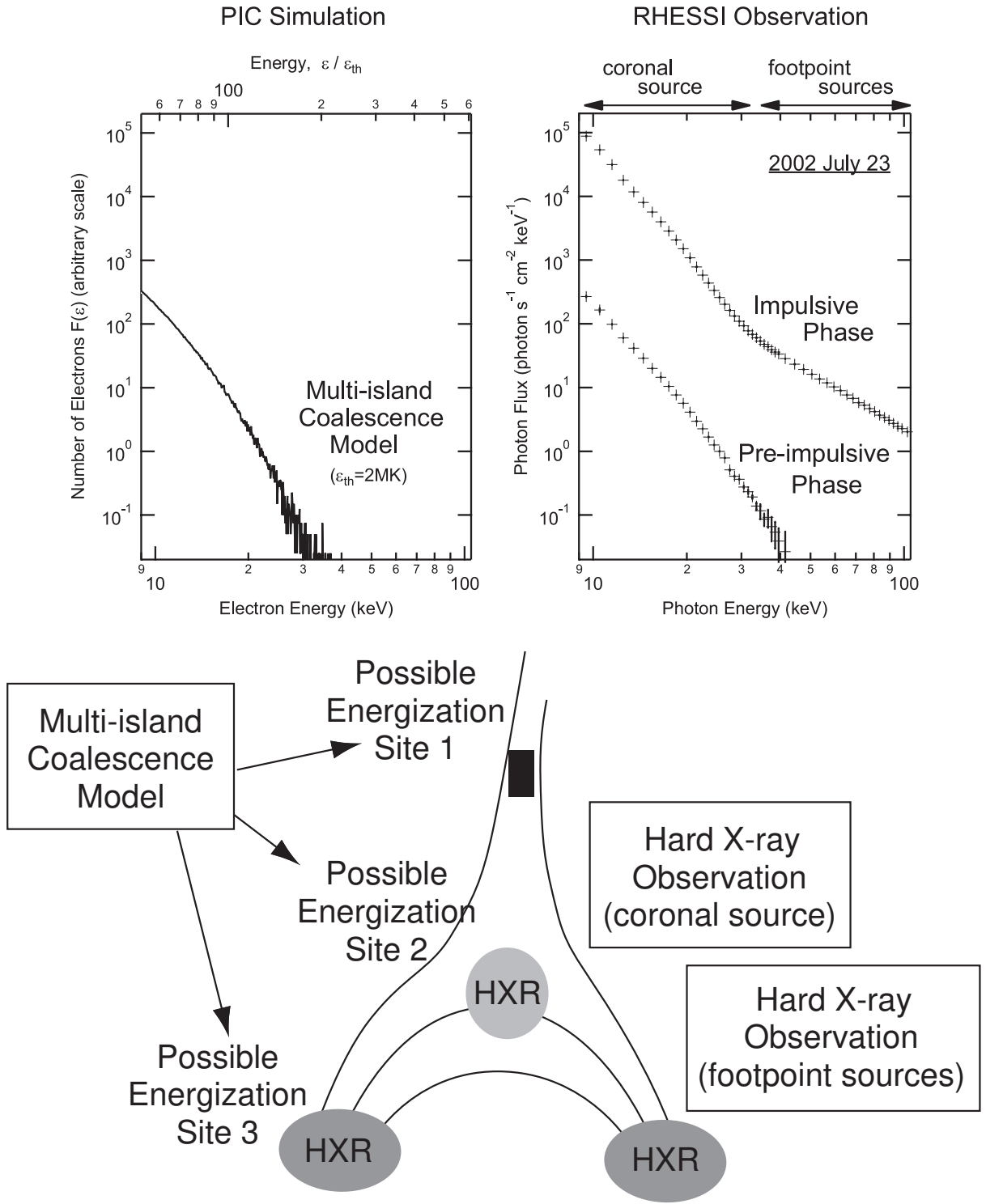}
\caption{(top left) Energy spectrum of electrons obtained by the PIC simulation. It is exactly the same as the one in Figure \ref{fig:spectrum_obs} but only shows the higher energy end.  The background temperature of 2MK is assumed to derive the electron energy. See texts for other assumptions and limitations of the simulation model. (top right) Energy spectra of X-ray photons obtained by the RHESSI satellite on 2002 July 23. The upper and the lower spectra are obtained at 0021:42 UT and 0028:10 UT, respectively. These are the reproduction of the energy spectra reported in Figure 2 of \cite{lin03} which paper contains the details of the event. (bottom) Schematic illustration of possible electron energization sites as well as typical coronal and footpoint sources of hard X-rays. 
\label{fig:rhessi}}
\end{center}
\end{figure*}

For the discussion, the energy spectrum in Figure \ref{fig:spectrum_obs} is revisited as shown in Figure \ref{fig:rhessi} (left).  A caveat here is that our simulation uses unrealistically high, initial temperature ($\varepsilon_{\rm th}$) due to limited computational resources.  The temperature of the initial background electrons $\varepsilon_{\rm th}$ = 9.216$\times$10$^{-3}$m$_{\rm e}c^2$ $\sim$ 55MK $\sim$5 keV which is already hot in the solar corona.  In Figure \ref{fig:rhessi} (left),  we re-normalized the energy by $\varepsilon_{\rm th}$ (top axis) and then converted to the unit of keV under the assumption of a typical electron temperature in the solar corona, $\varepsilon_{\rm th}$=2MK (bottom axis). Moreover, because the typical energy range of a space-borne, hard X-ray detector is from a few keV to a few hundreds of keV, we only showed the higher energy end of the simulated spectrum.  For the vertical axis, we used the number of electrons F($\varepsilon$) counted in the simulation box. F($\varepsilon$) can be representing a population $n$($\varepsilon$) keV$^{-1}$ cm$^{-3}$ of electrons in a fully ionized plasma of the energy release region. Since this is a simulation, the numbers can be scaled arbitrarily. 

As a reference, we reproduced in Figure \ref{fig:rhessi} (right) photon spectra from the well-known, solar flare event  of 23 July 2002 obtained by the RHESSI satellite \citep[e.g.][]{lin03,asai09}. In the pre-impulsive phase, the spectrum showed a power-law-like form that extends up to $>$40 keV from a coronal source. \cite{lin03} interpreted this spectrum by the combination of an exponential (thermal) and double power-law (non-thermal) spectra. In the impulsive phase, the energy range below 40 keV is dominated by a single exponential form, indicating clearly that electrons are significantly heated during this phase. Beyond this `super-hot' component is the power-law spectrum that extends up to $>$100 keV that originates from the chromospheric footpoints of the flare loop. 

It is worth emphasizing here that our simulation is not intended to reproduce any particular solar flare event. We must keep in mind that our simulation has many assumptions and limitations as described below. 
First, the spatial size of the simulation domain is very small $\sim$100 times the ion inertia length d$_{\rm i}$, that is $\sim$10$^4$cm or 0.14 milli-arcsec. This is much smaller than the typical size of the energy release region in solar flares, that is $\sim$10$^9$cm or 20 arcsec. Therefore, it is impossible to take into account the dynamical evolution of the solar flares. Second, the ion to electron mass ratio m$_{\rm i}$/m$_{\rm e}$ = 25 is very small compared to the actual value of 1836. Third, our simulation is a two-dimensional model in $x$ and $y$ so that spatial variations along the out-of-plane direction $z$ are not taken into account. Both reconnection and anti-reconnection creates strong current at and around the X-lines, and our simulation lacks possible consequences of current driven instabilities that would occur in the third dimension at the X-lines. Finally, we assumed zero magnitude of the magnetic field in the out-of-plane direction B$_{z}$, or the `guide field'. This assumption is merely for a simplicity. In order to better understand the physics of the solar flares, we must perform additional simulation runs with different guide field magnitude.

Nevertheless, the fact that our PIC simulation was able to produce electrons of up to $>$30 keV suggests that multi-island coalescence is potentially important for the understandings of electron acceleration/heating during the solar flares. Note that a multi-island coalescence may appear in association of developed turbulence that creates volume-filling magnetic island \citep{tajima97,  drake06, retino07, bemporad08}. Since turbulence may be generated in any possible electron energization site, our simulation may also be applied to these sites (Figure \ref{fig:rhessi}, bottom). The details of the possible electron energization sites as well as theories can be found elsewhere \citep[e.g.][and references therein]{aschwanden02}.  It is to be emphasized again that we used periodic boundaries but energetic particles that reached $\varepsilon>$1.4m$_{\rm e}$c$^2$ (or 26 keV when re-normalized) did not cross the boundary more than once (e.g. Figure \ref{fig:XL-overview} and Figure \ref{fig:OL-Drake1}), indicating that electrons do not need to move more than 100d$_{\rm i}$ (or $\sim$10$^4$ cm when re-normalized) in the $y$-direction to reach the energy. They also do not need to move more than 140d$_{\rm i}$ in the $z$-direction as we showed in Figure \ref{fig:diffusion}. Note also the total simulation time was $\sim$130$\Omega_{\rm ci}^{-1}$. If we assume the coronal magnetic field magnitude to be 100G, our simulation suggests that electrons of up to 30 keV can be created within a spatial extent of 10$^4$ cm within a time scale $\sim$ 10$\mu$sec. Such a localized and quick acceleration of electrons is an important feature of the multi-island coalescence.

Our PIC simulation also showed that the total energy gained by both thermal and non-thermal electrons reaches more than 30\% of the magnetic field energy released (Figure \ref{fig:overview_timeprofile}a). This relatively large number is the advantage of having multiple number of X-lines during the multi-island coalescence. Our simulation of single X-line reconnection with exactly the same parameters employed in the presented simulation showed the fraction of 20\% (not shown). Note that X-ray observations show that non-thermal electrons alone carry upwards of 10-50\% of the released magnetic field energy \citep[][and references therein]{lin03}. Therefore, our simulation may not fully explain the energy budget during the solar flares, but the multi-island coalescence is an attractive way to develop future models of the solar flare energy budget.



Figure \ref{fig:rhessi} suggests that the current PIC simulations produce similar spectra as observed in coronal hard X-ray sources, but cannot account for the flat electron spectrum needed to produce the hard X-ray spectrum observed in footpoints. In fact, if we were able to perform PIC simulation with a larger simulation domain, say 1000d$_{\rm i} \times$ 1000d$_{\rm i}$, the third spectral component described in Figure \ref{fig:spectrum_obs} may extend to higher energy range to form a clear power-law. Moreover, there would be a variety of different size of magnetic islands so that the resultant turbulence leads to a Fermi-like, stochastic acceleration of electrons.  Therefore,
 we anticipate our results to be a starting point for more sophisticated models of particle acceleration during the explosive energy release phenomena.

\section{Conclusion}

We performed 2D PIC simulation to study electron acceleration during multi-island coalescence. By analyzing the trajectories of the most energetic electrons, we found a variety of different acceleration mechanisms such as the contracting island mechanism, ripple mechanism, island surfing mechanism, etc. However, a statistical study showed that the most important process is the energization process that takes place at the anti-reconnection region. Based on the maximum energy of electrons attained in the simulation,  we pointed out that the multi-island coalescence may play an important role in producing energetic electrons during the solar flares.

\acknowledgments


We acknowledge helpful discussion with M. Shay, S. Zenitani, H. Hudson, P. L. Pritchett, H. Isobe, S. Imada, M. Hirai and F. Mozer. Simulations are performed on the SX-9 computer of JAXA, Japan. We thank all members of the RHESSI project for providing the hard X-ray data. This research was funded by NSF grant ATM-0613886 and NASA grant NNX08AO83G at UC Berkeley. This work was M.O. acknowledges additional support from the Grant-in-Aid for Creative Scientific Research (17GS0208) from the MEXT, Japan.

\clearpage

\clearpage


\begin{thebibliography}{}
\bibitem[Ambrosiano et al.(1988)]{ambrosiano88} Ambrosiano, J. et al. 1988, \jgr, 93, 14383
\bibitem[Asai et al.(2009)]{asai09} Asai, A. et al. 2009, \apj, 695, 1623
\bibitem[Aschwanden (2002)]{aschwanden02} Aschwanden, M. 2002, {\it Space Sci. Rev.}, 101, 1
\bibitem[Bemporad (2008)]{bemporad08} Bemporad, A. 2008, \apj, 689, 572.
\bibitem[Bemporad et al.(2006)]{bemporad06} Bemporad, A. et al. 2006, \apj, 638, 1110
\bibitem[Benz (2008)]{benz08} Benz, A., Living Reviews in Solar Physics, 5, 1
\bibitem[Biskamp \& Welter (1980)]{biskamp80} Biskamp, D. \& Welter, H., 1980, \prl, 44, 1069
\bibitem[Daughton et al.(2006)]{daughton06} Daughton, W., Scudder, J. \& Karimabadi, H. 2006, Phys. Plasmas, 13, 072101
\bibitem[Drake et al.(2005)]{drake05} Drake, J. F. et al. 2005, \prl, 94, 095001
\bibitem[Drake et al.(2006)]{drake06} Drake, J. F. et al. 2006, Nature, 443, 05116
\bibitem[Finn \& Kaw(1977)]{finn77} Finn, J. M. \& Kaw, P. K. 1977, Phys. Fluids 20, 72
\bibitem[Fujimoto (2006)]{fujimoto06} Fujimoto, K. 2006, Phys. Plasmas 13, 072904
\bibitem[Hannah \& Fletcher(2006)]{hannah06} Hannah, I. \& Fletcher, L. 2006, Solar Physics, 236, 59
\bibitem[Hoshino, 1987]{hoshino87} Hoshino, M. 1987, \jgr, 92, 7368
\bibitem[Hoshino et al. (2001)]{hoshino01} Hoshino, M. et al. 2001, \jgr, 106, 25979
\bibitem[Hoshino (2005)]{hoshino05} Hoshino, M. 2005, \jgr, 110, A10215
\bibitem[Karimabadi et al. (2007)]{karimabadi07} Karimabadi, H., Daughton, W. \& Scudder, J. 2007, \grl, 34, L13104
\bibitem[Karlick\'{y} \& B\'{a}rta (2007)]{karlicky07} Karlick\'{y}, M. \& B\'{a}rta, M. 2007, A\&A, 464, 735.
\bibitem[Kliem (1994)]{kliem94} Kliem, B. 1994, \apj, 90, 719
\bibitem[Krucker et al.(2008)]{krucker08} Krucker, S. et al. 2008, {\it Astron. Astrophys. Rev.}, 16, 155.
\bibitem[Lin et al.(2003)]{lin03} Lin, R. P. et al. 2003, \apj, 595, L69
\bibitem[Litvinenko (1996)]{litvinenko96} Litvinenko, Y. E. 1996, \apj, 462, 997
\bibitem[Miller et al.(1997)]{miller97} Miller, J. A. et al, \jgr, 102, 14631
\bibitem[Oka et al.(2010)]{oka10b} Oka, M. et al., 2010, submitted to \jgr
\bibitem[Onofri, Isliker \& Vlahos (2006)]{onofri06} Onofri, I., Isliker, H. \& Vlahos, L. 2006, \prl, 96, 151102
\bibitem[Phan et al.(2007)]{phan07} Phan, T. D. et al. 2007, \prl, 99, 255002
\bibitem[Pritchett \& Wu (1979)]{pritchett79} Pritchett, P. L. \& Wu, C. C. 1979, Phys. Fluids, 22, 2140
\bibitem[Pritchett (1980)]{pritchett80} Pritchett, P. L. et al. 1980, Phys. Fluids 23(7), 1368
\bibitem[Pritchett (2006)]{pritchett06} Pritchett, P. L. 2006, Geophys. Res. Lett., 33, L13104
\bibitem[Pritchett (2007)]{pritchett07} Pritchett, P. L. 2007, Phys. Plasmas 14, 052102
\bibitem[Pritchett (2008)]{pritchett08} Pritchett, P. L. 2008, Phys. Plasmas 15, 102105
\bibitem[Retin\`{o} et al.(2007)]{retino07} Retin\`{o}, A. et al. 2007, Nature Physics 3, 236
\bibitem[Sato et al. (1982)]{sato82} Sato, T., Matsumoto, H., \& Nagai, K. 1982, \jgr, 87, 6089
\bibitem[Scholer and Jamitzky (1987)]{scholer87} Scholer, M. \& Jamitzky, F. 1987, \jgr, 92, 12181
\bibitem[Shay et al.(2007)]{shay07} Shay, M. A., Drake, J. F., \& Swisdak, M. 2007, \prl, 99, 155002
\bibitem[Shinohara et al.(2001)]{shinohara01} Shinohara, I. et al. 2001, \prl,  87, 095001
\bibitem[Shinohara et al.(2009)]{shinohara09} Shinohara, I. et al. 2009, {\it Future Perspectives of Space Plasma and Particle Instrumentation and International Collaborations}, pp.248
\bibitem[Stern (1979)]{stern79} Stern, D. P. 1979, \jgr, 84, 63
\bibitem[Sui \& Holman (2003)]{sui03} Sui, L. \& Holman, G. D. 2003, \apj, 596, L251
\bibitem[Tajima et al. (1987)]{tajima87} Tajima, T. et al. 1987, \apj, 321, 1031
\bibitem[Tajima \& Shibata (1997)]{tajima97} Tajima, T. \& Shibata, K. 1997, {\it Plasma Astrophysics} (Massachusetts: Addison-Wesley)
\bibitem[Tanaka et al.(2010)]{tanaka10} Tanaka, K. et al., submitted to \prl
\bibitem[Wan et al. (2008)]{wan08} Wan, W. et al. 2008, Phys. Plasmas 15, 032903
\bibitem[Wan \& Lapenta (2008)]{wan08c} Wan, W. \& Lapenta, G. 2008, \prl 100, 035004

\end{thebibliography}
\end{document}